\documentclass[prd,nofootinbib,twocolumn,superscriptaddress,preprintnumbers,balancelastpage]{revtex4-1}

\pdfoutput=1
\usepackage{amsmath,amssymb}
\usepackage{amsfonts}
\usepackage{bm,bbm}
\usepackage{graphicx}
\usepackage{mathrsfs}
\usepackage{slashed}
\usepackage{booktabs}
\usepackage{tabu}
\usepackage[dvipsnames]{xcolor}
\usepackage[normalem]{ulem}
\usepackage{tikz}
\usepackage{soul}

\usepackage{hyperref}
\hypersetup{colorlinks,citecolor= blue,linkcolor= blue, urlcolor=blue}

\usepackage{xcolor}
\usepackage{ulem}
\usepackage{array}
\usepackage{verbatim}
\usepackage{epsfig}
\usepackage{multirow}

\newcommand{\be}{\begin{equation}}
\newcommand{\ee}{\end{equation}}
\newcommand{\ba}{\begin{array}}
\newcommand{\ea}{\end{array}}
\newcommand{\bea}{\begin{eqnarray}}
\newcommand{\eea}{\end{eqnarray}}

\usepackage{ulem,fancyvrb}
\usepackage{xcolor}

\definecolor{blue-violet}{rgb}{0.54, 0.17, 0.89}
\definecolor{amethyst}{rgb}{0.6, 0.4, 0.8}

\begin{document}

\title{Supernova constraints on lepton flavor violating axions}

\author{Yonglin Li}%\email{yonglinli@smail.nju.edu.cn}
\affiliation{Department of Physics, Nanjing University, Nanjing 210093, China}
\author{Zuowei Liu}%\email{zuoweiliu@nju.edu.cn}
%\thanks{Corresponding author.\\ zuoweiliu@nju.edu.cn}
\affiliation{Department of Physics, Nanjing University, Nanjing 210093, China}

\begin{abstract}

Supernovae offer a unique hot and dense environment to probe new physics beyond the Standard Model. We investigate supernova cooling constraints on lepton-flavor-violating (LFV) axions and axion-like particles (ALPs) that couple to electrons and muons. For LFV-ALP production in supernovae, muon decay and lepton bremsstrahlung have been considered previously. In this work, we identify the electron-muon coalescence channel as an efficient new production mechanism in the high-mass regime. We also include the semi-Compton scattering process, which has recently been shown to provide sizable contributions for electron-coupled ALPs. We find that muon decay dominates in the low-mass regime, electron-muon coalescence becomes the leading channel at high masses, and semi-Compton scattering provides the dominant contribution in the intermediate mass range. We find that the electron-muon coalescence process yields the strongest constraints in the mass range of $\sim (115,280)$ MeV, probing the ALP-electron-muon coupling down to $\sim 4\times 10^{-10}$ for an ALP mass of $\sim200$ MeV. 

\end{abstract}

\maketitle

\section{Introduction}

Axions
are beyond the standard model (BSM) pseudo-scalar particles 
that were originally proposed to solve the strong CP problem  
\cite{Peccei:1977hh,Wilczek:1977pj,Weinberg:1977ma,Peccei:2006as}.
In addition to 
axions, a number of hypothetical pseudo-scalar particles, 
referred to as axion-like particles (ALPs), 
can arise in a variety of well-motivated BSM models 
\cite{Gelmini:1980re,Davidson:1981zd,Wilczek:1982rv,Svrcek:2006yi}. 
ALPs can couple to various standard model (SM) particles 
in ways that lead to violations of SM symmetries, 
such as lepton flavor conservation \cite{Wilczek:1982rv}.  
The lepton-flavor-violating (LFV) ALPs can arise 
either from UV theories of ALPs 
\cite{Davidson:1981zd,Wilczek:1982rv,Anselm:1985bp,Feng:1997tn,Bauer:2016rxs,Ema:2016ops,Calibbi:2016hwq}, 
which are known as flavons or familons,   
or from radiative corrections 
\cite{Choi:2017gpf,Chala:2020wvs,Bauer:2020jbp}, 
even if the underlying theory preserves flavor.

Thus, searching for LFV-ALPs is an intriguing   
aspect of new physics studies
beyond the SM.  
In particular, 
accelerator experiments provide the best constraints 
on heavy LFV-ALPs with mass above GeV 
\cite{Endo:2020mev,Iguro:2020rby,
Davoudiasl:2021haa,Cheung:2021mol,Davoudiasl:2021mjy,
Araki:2022xqp,Calibbi:2022izs,Batell:2024cdl,Calibbi:2024rcm}. 
In contrast, sub-GeV LFV-ALPs 
are best probed in decay experiments \cite{Derenzo:1969za,
Jodidio:1986mz,Bryman:1986wn,Bilger:1998rp,TWIST:2014ymv,
Bauer:2019gfk,Cornella:2019uxs,PIENU:2020loi,Calibbi:2020jvd,
Bauer:2021mvw,Jho:2022snj,Knapen:2023zgi,Knapen:2024fvh},
and in stellar processes \cite{Calibbi:2020jvd,Zhang:2023vva}. 
Among the stellar objects, core-collapsed supernovae (CCSNe), 
the astrophysical events 
that provide extremely dense and hot environments, 
probe LFV-ALPs with a relatively high mass up to 100 MeV
\cite{Calibbi:2020jvd}. 
The most renowned example of CCSNe is the SN 1987A, 
whose neutrino signals not only provide crucial 
insights into the collapse and explosion mechanisms of such supernovae, 
but also place stringent constraints on potential new energy-loss channels 
induced by BSM particles, 
which is known as the SN cooling limit 
\cite{Raffelt:1996wa}.

Recently, 
the SN cooling limits on LFV-ALPs 
that couple both to electrons and muons 
have been analyzed by 
Refs.~\cite{Calibbi:2020jvd,Zhang:2023vva}. 
To compute the axion production rate in the SN, 
Ref.~\cite{Calibbi:2020jvd} considered  
the muon decay process, $\mu\to e+a$, 
where $a$ is the ALP, and  
Ref.~\cite{Zhang:2023vva} 
considered the lepton bremsstrahlung process. 
In this paper, we further consider the 
electron-muon coalescence process, 
$e^\mp + \mu^\pm \to a$, 
for the ALP production in the SN. 
We find that for heavy ALPs with mass of
$m_a>m_e+m_\mu$, 
the electron-muon coalescence process is the 
dominant ALP production channel in the SN core for 
the parameter space of interest. 
Recently, Ref.~\cite{Fiorillo:2025sln}
showed that the semi-Compton 
process can provide significant contributions 
for ALPs that couple to electrons.  
Thus, in this work, we also consider 
the semi-Compton process,   
$e/\mu+\gamma\to \mu/e + a$, 
for LFV-ALP production in the SN. 
We find that the semi-Compton process 
provides the leading contribution 
in the intermediate mass range, 
dominating over the lepton bremsstrahlung process 
in that region.

In computing SN cooling limits, 
Refs.~\cite{Calibbi:2020jvd,Zhang:2023vva} 
assumed an SN core with constant density and temperature. 
In this work, 
we use profiles from recent supernova simulations 
\cite{Bollig:2020xdr,garching-profile}, 
which 
offer a more accurate description of the SN core. 
These include profiles of temperature and 
chemical potentials of various particles, 
as shown in Fig.~(\ref{fig:SN-profile}). 
Moreover, Refs.~\cite{Calibbi:2020jvd,Zhang:2023vva} 
neglected the effects of ALP absorption.
Here, we compute the ALP luminosity by properly taking 
into account the ALP absorption effects, 
following Refs.~\cite{Caputo:2021rux,Caputo:2022rca}; 
see also e.g., Refs.~\cite{Lucente:2022vuo,
Carenza:2023lci,Carenza:2023old}. 
We find that the ALP absorption effects are 
significant and 
must be accounted for in order to obtain more accurate constraints.

In this paper, 
we compute the ALP luminosity 
by considering the following 
four production channels in the SN core:
(i) muon decay, 
(ii) lepton bremsstrahlung, 
(iii) electron-muon coalescence, and 
(iv) semi-Compton. 
By using recent SN profiles and 
taking into account the absorption effects, 
we find that the SN cooling constraints on 
LFV-ALPs probe 
currently unexplored parameter space in the mass 
range of 
$\sim (105,280)$ MeV; 
where 
the dominant ALP production channels are the semi-Compton process 
for the mass range of $\sim(105,115)$ MeV 
and the electron-muon coalescence process 
for $\sim(115,280)$ MeV.
Note that in this mass range, 
LFV-ALPs cannot be probed by 
the rare muon decay experiments due to the 
kinematical conditions. 
Collider experiments, on the other hand, 
only probe the large coupling regime 
in this mass range, 
which does not overlap with the parameter space probed by SN processes; 
see e.g., Refs.~\cite{Cornella:2019uxs,Endo:2020mev}. 
Thus, SN cooling limits arising from  
the electron-muon coalescence process 
offer a unique probe of the high ALP mass range.

The rest of the paper is organized as follows. 
In section \ref{sec:ALP-model} we discuss  
the LFV-ALP model. 
In section \ref{sec:SN-model} we discuss 
the SN profiles 
and the ALP luminosity 
in the SN cooling analysis. 
In section \ref{sec:ALP-prod-abs} we discuss the 
ALP production and absorption rates. 
Our results are given in 
section \ref{sec:result}. 
We summarize our findings in \ref{sec:sum}. 
We further provide detailed calculations of 
the ALP production rate in 
appendix \ref{sec:ALP-prod}, 
and of the ALP absorption 
rate in appendix \ref{sec:abs}.

\section{ALP Model}
\label{sec:ALP-model}

We consider an ALP model 
where the electron number and the muon number are violated 
by the interaction with ALPs: 
\begin{equation}
    \mathcal{L}_{\rm int} = \frac{g_{ae\mu}}{m_e^0+m_\mu}
    \bar e \gamma^\lambda \gamma_5 \mu 
    \partial_\lambda a + {\rm h.c.},
    \label{eq:lagrangian}
\end{equation}
where $g_{ae\mu}$ is the coupling constant, 
$a$ is the ALP field, 
$e$ ($\mu$) 
is the electron (muon) field, 
and $m_e^0=0.511$ MeV ($m_\mu=105.6$ MeV) 
is the electron (muon) mass. 
Note that we use $m_e^0$ to denote the electron mass 
in the vacuum, which should be distinguished 
from the in-medium 
electron mass in the SN core, $m_e$, 
as it 
receives significant corrections due to plasma effects. 
At tree level the interaction Lagrangian given in 
Eq.~\eqref{eq:lagrangian} 
is equivalent to \cite{Raffelt:1987yt}
\begin{equation}
    \mathcal{L}_{\rm int} = -ig_{ae\mu} a 
    \bar e \gamma_5 \mu + {\rm h.c.}. 
    \label{eq:eff-lagrangian}
\end{equation}
We note that 
Eq.~\eqref{eq:eff-lagrangian} is equivalent to Eq.~\eqref{eq:lagrangian} when 
both fermions are on-shell, 
which can be proven using the equations of motion; 
see also Refs.~\cite{Lucente:2021hbp,Ferreira:2022xlw}. 
However, when there exist off-shell fermions, 
the equivalence between 
Eq.~\eqref{eq:eff-lagrangian} and Eq.~\eqref{eq:lagrangian} 
is lost. 
In our analysis we use the Lagrangian given in 
Eq.~\eqref{eq:eff-lagrangian} to 
compute the SN limits.

\section{SN Model}
\label{sec:SN-model}

\begin{figure}[htbp]
\centering
\includegraphics[width=0.5\textwidth]{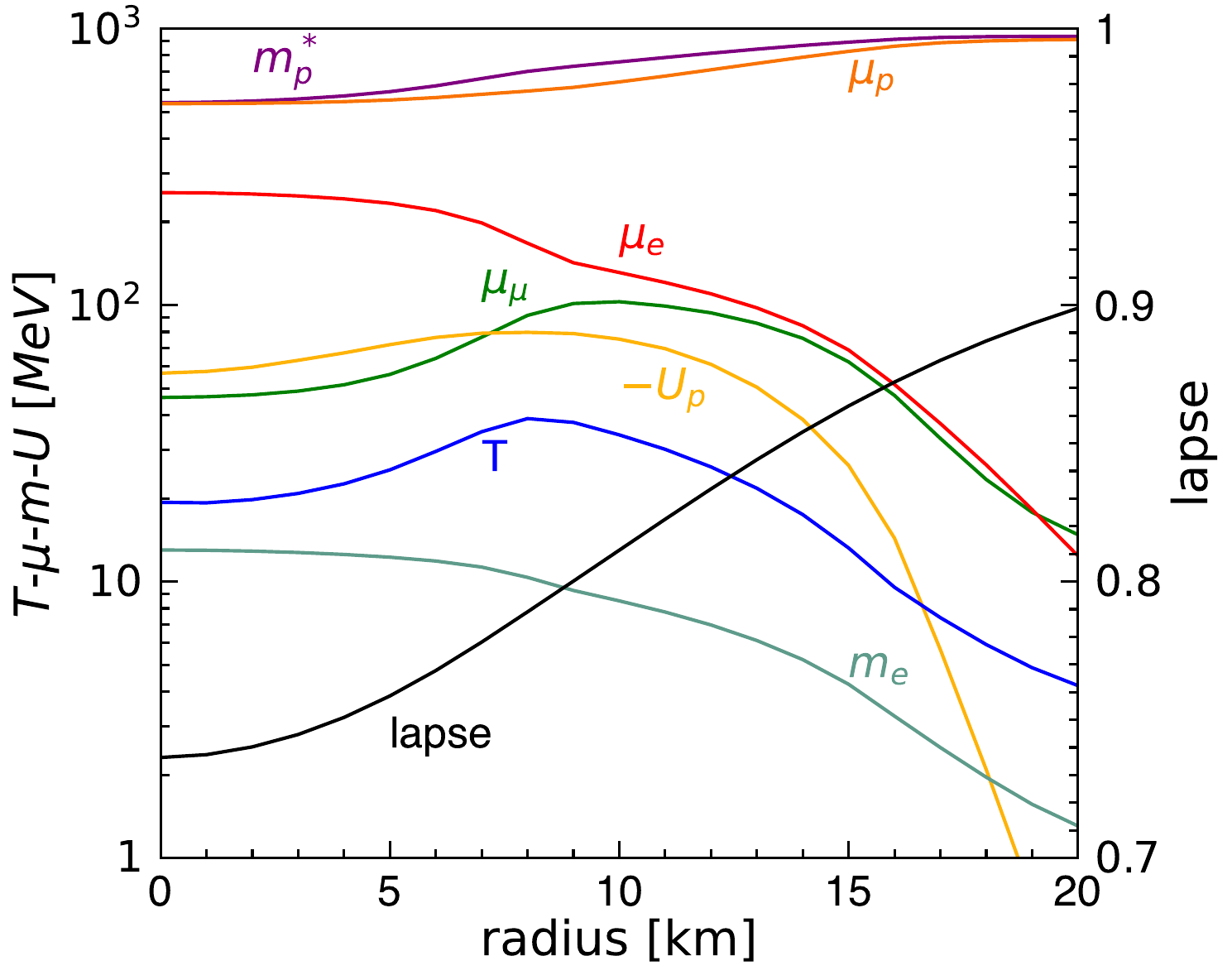}
\caption{The profiles of temperature $T$, 
electron chemical potential $\mu_e$, 
muon chemical potential $\mu_\mu$, 
proton chemical potential $\mu_p$, 
effective proton mass $m_p^*$, and 
the negative of the proton interaction potential 
$U_p$, as well as the gravitational lapse,
for the Garching muonic SN model SFHo-18.8
at 1 second postbounce 
\cite{garching-profile}. 
The effective electron mass is calculated with 
$T$ and $\mu_e$ using Eq.~\eqref{eq:e-mass}. 
Note that the proton chemical potential 
$\mu_p$
given in 
\cite{garching-profile} 
excludes the rest mass; 
here, we include the rest mass in $\mu_p$.
}
\label{fig:SN-profile}
\end{figure}

The production rate of the LFV-ALP 
depends strongly on the temperature $T$, 
and on the number densities of various particles, 
which in turn depend on their chemical potentials. 
In our analysis, 
we adopt the temperature and chemical potential profiles 
from recent simulations 
in the Garching muonic SN model SFHo-18.8 
\cite{Bollig:2020xdr,garching-profile}. 
Among the models simulated in Refs.~\cite{Bollig:2020xdr,garching-profile}, 
SFHo-18.8 is the coldest and thus provides 
a conservative and robust constraint on LFV-ALPs. 
Fig.~(\ref{fig:SN-profile}) shows various 
profiles at 1 second postbounce, including temperature $T$, 
electron chemical potential $\mu_e$, 
muon chemical potential $\mu_\mu$, 
and proton chemical potential $\mu_p$.

We note that 
muons reach thermal equilibrium with other SM particles via electromagnetic interactions, 
but their chemical potential is determined by weak processes that change the muon number. 
In the context of SN cooling, the ALP processes considered here 
alter the muon number by one unit and are required to be comparable in strength to weak interactions. 
This may lead to non-negligible modifications to the muon distribution in the SN core. 
These effects, however, are neglected in our analysis.

In a plasma environment like the SN core, 
the properties of electrons are modified significantly so that
the electron mass in the vacuum ($m_e^0=0.511$ MeV) is 
replaced by the in-medium electron mass  
\cite{Braaten:1991hg,Lucente:2021hbp}: 
\begin{equation}
    m_e = \frac{m_e^0}{\sqrt{2}}+\sqrt{\frac{(m_e^0)^2}{2}
    +\frac{\alpha}{\pi}(\mu_e^2+\pi^2T^2)}, 
    \label{eq:e-mass}
\end{equation}
where $\alpha=1/137$ is the fine-structure constant. 
We compute $m_e$ 
as a function of the radius $r$ by using 
the $\mu_e$ and $T$ profiles. 
As shown in Fig.~(\ref{fig:SN-profile}), 
the in-medium electron mass 
is in the range of 
$8.3$ MeV $\lesssim m_e \lesssim 12.9$ MeV 
for $r\lesssim 10$ km. 

For the lepton bremsstrahlung process, the ALP production rate 
depends on the profile of the proton number density, 
which in turn depends on the 
profiles of the effective proton mass $m_p^*$,
the proton chemical potential $\mu_p$,
and the proton interaction potential 
$U_p$, 
which are shown in Fig.~(\ref{fig:SN-profile}); 
see appendix \ref{sec:ALP-prod} 
for detailed calculations.
The ALP luminosity also depends on 
the lapse factor, 
which accounts for both the gravitational redshift 
as well as contributions 
from the pressure and energy of both the stellar medium 
and neutrinos \cite{1979ApJ...232..558V,Shapiro:1983du,Rampp:2002bq,Caputo:2022mah}. 
The profile of the lapse factor 
is shown in Fig.~(\ref{fig:SN-profile}).

\subsection{ALP luminosity}
\label{sec:ALP-lum}

ALPs can introduce a new cooling channel of the SN core, 
which can significantly modify the properties of the SN  
and thus affects its neutrino luminosity \cite{Raffelt:1996wa}.
Based on the observation of SN 1987A \cite{Kamiokande-II:1987idp}, 
Ref.~\cite{Raffelt:1996wa} established the criterion 
that the luminosity of the ALPs from the SN inner core 
should be smaller than the luminosity of neutrinos 
at about 1 second postbounce: 
\begin{equation}
    L_a\leq L_\nu = 3\times 10^{52}\ {\rm erg/s}. 
\end{equation}
This is known as the supernova cooling limit.
If the couplings between ALPs and SM particles are weak, 
ALPs can free-stream in the SN, and
their luminosity 
can be computed via simple volume integrations,
neglecting reabsorption effects.
However, if the couplings are 
strong, ALPs may be rapidly reabsorbed inside the SN core, 
significantly altering their escape and resulting luminosity.

\begin{figure*}[htbp]
\centering
\includegraphics[width=0.3 \textwidth]{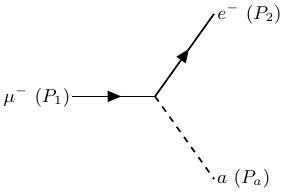}
\hspace{0.1cm}
\includegraphics[width=0.6 \textwidth]{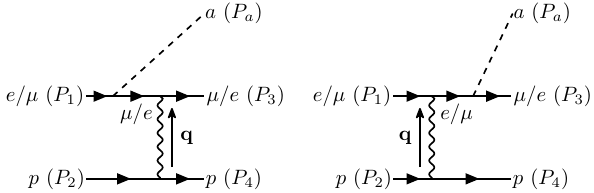}
\\[0.5 cm]
\includegraphics[width=0.28 \textwidth]{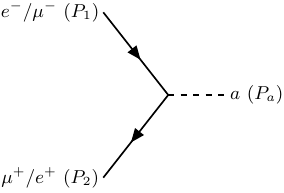}
\hspace{0.1cm}
\includegraphics[width=0.6\textwidth]{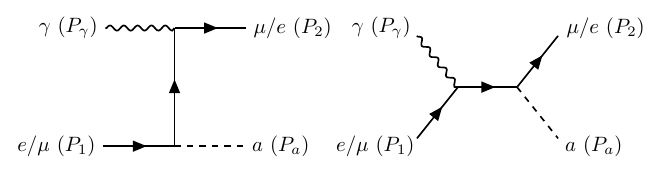}
\caption{
LFV-ALP production processes in the SN. Upper: muon decay (left) 
and lepton bremsstrahlung (middle and right). Lower: $e$-$\mu$ 
coalescence (left) and semi-Compton (middle and right).
}
\label{fig:feynman}
\end{figure*}

The SN cooling limits on LFV-ALPs 
have been studied for the muon decay process \cite{Calibbi:2020jvd} 
and for the lepton bremsstrahlung process \cite{Zhang:2023vva}. 
Here, we further consider 
the $e$-$\mu$ coalescence process 
and the semi-Compton process for ALP production. 
Thus, we consider the following four processes: 
(i) muon decay, 
(ii) lepton bremsstrahlung, 
(iii) $e$-$\mu$ coalescence, and
(iv) semi-Compton. 
The Feynman diagrams of the four processes 
are shown in Fig.~(\ref{fig:feynman}).
We find that the $e$-$\mu$ coalescence and 
the semi-Compton processes, 
which were neglected in 
Refs.~\cite{Calibbi:2020jvd,Zhang:2023vva},
lead to new constraints on LFV-ALPs  
in the mass range of $(105,280)$ MeV.

To compute the SN cooling limit, we compute 
the ALP luminosity at the gain radius $R_g$ 
\cite{Caputo:2022rca} 
\cite{Caputo:2022mah}: 
\begin{align}
    L_a =& \int_0^{R_g} dr\ 
    4\pi r^2\, {\rm lapse}(r)^2(1+2v_r) \nonumber  \\
    &\times\int_{m_a'}^\infty dE_a\ 
    E_a\frac{d^2n_a}{dt dE_a}\langle e^{-\tau_a(R_g,E_a,r)}\rangle,
    \label{eq:lum-obs}
\end{align}
where $E_a$ is the ALP energy, 
$n_a$ is the ALP number density, 
$\tau_a$ is the optical depth,  
$\langle e^{-\tau_a(R_g,E_a,r)}\rangle$ accounts for 
the absorption of the ALP, 
${\rm lapse}(r)$ is the lapse factor 
accounting for the 
gravitational redshift effects, 
$v_r$ is the radial velocity of 
the emitting medium, 
and $m_a' = m_a/{\rm lapse(r)}$. 
In our analysis we neglect $v_r$, since $|v_r|\ll 1$ 
\cite{Caputo:2022mah}. 
We note that the absorption effects of ALPs were    
neglected in previous studies on SN cooling constraints 
on LFV-ALPs \cite{Calibbi:2020jvd,Zhang:2023vva}. 
However, we find that the absorption effects, 
which are properly taken into account in Eq.~\eqref{eq:lum-obs}, 
play a significant role in the ALP luminosity.  
We also note that $L_a$ computed via Eq.~\eqref{eq:lum-obs} 
can be directly compared to the neutrino luminosity at 
infinity, as the gravitational redshift effects have been taken 
into account. 
We refer to ${d^2n_a}/{dt dE_a}$, 
the number of ALPs 
produced per unit volume per unit time per unit energy, 
as  ``the ALP production rate'' hereafter. 
In our analysis we use $R_g=21.0$ km, 
which is the gain radius at 1 second 
postbounce in Ref.~\cite{garching-profile}.

The absorption term is computed via 
\cite{Caputo:2021rux,Caputo:2022rca}
\begin{equation}
    \langle e^{-\tau_a(R_g,E_a,r)}\rangle 
    = \int_{-1}^1  \frac{d\mu}{2}
    e^{-\int_0^{s_{\rm max}} \frac{ds}{v} \Gamma_{\rm abs}
    (E_a,\sqrt{r^2+s^2+2rs\mu})},
    \label{eq:abs-full}
\end{equation}
where $v$ is the velocity of the ALP, 
$\Gamma_{\rm abs}$ is the absorption rate of the ALP, 
$\mu=\cos\beta$ with $\beta$ the angle between 
outward radial direction and the trajectory 
of the ALP along which $ds$ is integrated, 
and $s_{\rm max} = \sqrt{R_g^2-r^2(1-\mu^2)}-r\mu$ 
is upper limit of the $ds$ integration. 
Note that the absorption 
up to the gain radius $R_g$ is taken into account 
such that $s_{\rm max}$ satisfies 
$r^2+s_{\rm max}^2+2r s_{\rm max}\mu = R_g^2$.

We note that different treatments 
on the new particle luminosity have been previously used in the literature. 
For example, 
Ref.~\cite{Chang:2016ntp} considered new particle production 
up to $R_\nu$ and computed the absorption up to $R_g$; 
see also Refs.~\cite{Lucente:2020whw,
Lucente:2021hbp,Ferreira:2022xlw}. 
In contrast, 
we consider ALP production within 
$R_g$ as in Eq.~\eqref{eq:lum-obs}, 
rather than using $R_\nu$
as in Refs.~\cite{Chang:2016ntp,Lucente:2020whw,
Lucente:2021hbp,Ferreira:2022xlw}. 
This distinction is motivated by the fact that the energy within $R_g$ 
contributes to neutrino production \cite{Chang:2016ntp},
and 
ALP production in the region between $R_\nu$ and $R_g$ also 
depletes the energy that is available for neutrino emission.  
Therefore, the ALP production between $R_\nu$ and $R_g$ 
should be taken into account when computing the SN cooling constraints. 
We note that Ref.~\cite{Caputo:2022rca} has computed the ALP luminosity 
consistently at any radius.

\section{ALP production and absorption}
\label{sec:ALP-prod-abs}

To compute the ALP production rate 
${d^2n_a}/{dtdE_a}$ in Eq.~\eqref{eq:lum-obs},
we consider four ALP production channels in the SN core:  
(i) muon decay, 
(ii) lepton bremsstrahlung, 
(iii) electron-muon coalescence, and 
(iv) semi-Compton. 
For light ALPs with $m_a<m_\mu-m_e$, 
the muon decay process 
$\mu^-\to e^-+a$ is kinematically allowed, 
as shown in the upper-left diagram of 
Fig.~(\ref{fig:feynman}). 
For heavy ALPs with $m_a>m_\mu+m_e$, the electron-muon 
coalescence process $e^\mp + \mu^\pm \to a$ 
becomes kinematically accessible, 
as shown in the lower-left diagram of Fig.~(\ref{fig:feynman}). 
The lepton bremsstrahlung process, as shown in the 
upper-middle and upper-right diagrams of 
Fig.~(\ref{fig:feynman}), contains two sub-processes: 
$e^-p\to \mu^- pa$ and $\mu^-p\to e^- pa$. 
The semi-Compton process contains two sub-processes: 
$e^- + \gamma \to \mu^- + a$ and $\mu^- + \gamma \to e^- + a$, 
as shown in the lower-middle 
and lower-right diagrams of Fig.~(\ref{fig:feynman}). 
Because the lepton bremsstrahlung and 
semi-Compton processes are kinematically allowed 
for any ALP mass, they are the only 
viable ALP production channels
in the mass range of $-m_e < m_a-m_\mu < m_e$, 
as the other two processes are 
kinematically forbidden.
We find that the production rate 
for the $e^-+\mu^+\to a$ channel is higher than 
that for $e^+ +\mu^-\to a$ in the SN. 
We note that the electron-muon 
coalescence process 
and the semi-Compton process have 
not been considered 
in previous studies 
on LFV-ALPs in SNe \cite{Calibbi:2020jvd,Zhang:2023vva}. 
We find that, 
despite the relatively low abundances of 
positrons and anti-muons, 
the electron-muon 
coalescence process dominates ALP production 
in SNe for the mass range of $m_a>m_\mu+m_e$.

To compute the ALP absorption rate $\Gamma_{\rm abs}$ in Eq.~\eqref{eq:abs-full}, 
we consider the following four processes: 
(i) $e$-$a$ coalescence $e^-+a\to \mu^-$,
(ii) inverse bremsstrahlung, 
(iii) ALP decay $a\to e^\pm + \mu^\mp$, and
(iv) inverse-Compton, 
which 
are the inverse of the 
four ALP production processes shown in 
Fig.~(\ref{fig:feynman}). 
See appendix \ref{sec:ALP-prod} and \ref{sec:abs} 
for the detailed calculations on 
the ALP production rates and 
absorption rates.

We note that for light ALPs, the muon decay process 
dominates the ALP production in the SN. 
In this regime, 
both the production and absorption rates  
are largely insensitive to the ALP mass, 
and the absorption effects 
are negligible for $g\lesssim 10^{-9}$. 
In contrast, 
for heavier ALPs where the electron-muon coalescence 
process dominates the ALP production, 
the production rate and 
absorption rate become sensitive to both 
the coupling constant $g$ and the ALP mass $m_a$.

\section{Results and discussions}
\label{sec:result}

\begin{figure}[htbp]
    \centering
    \includegraphics[width=0.45\textwidth]{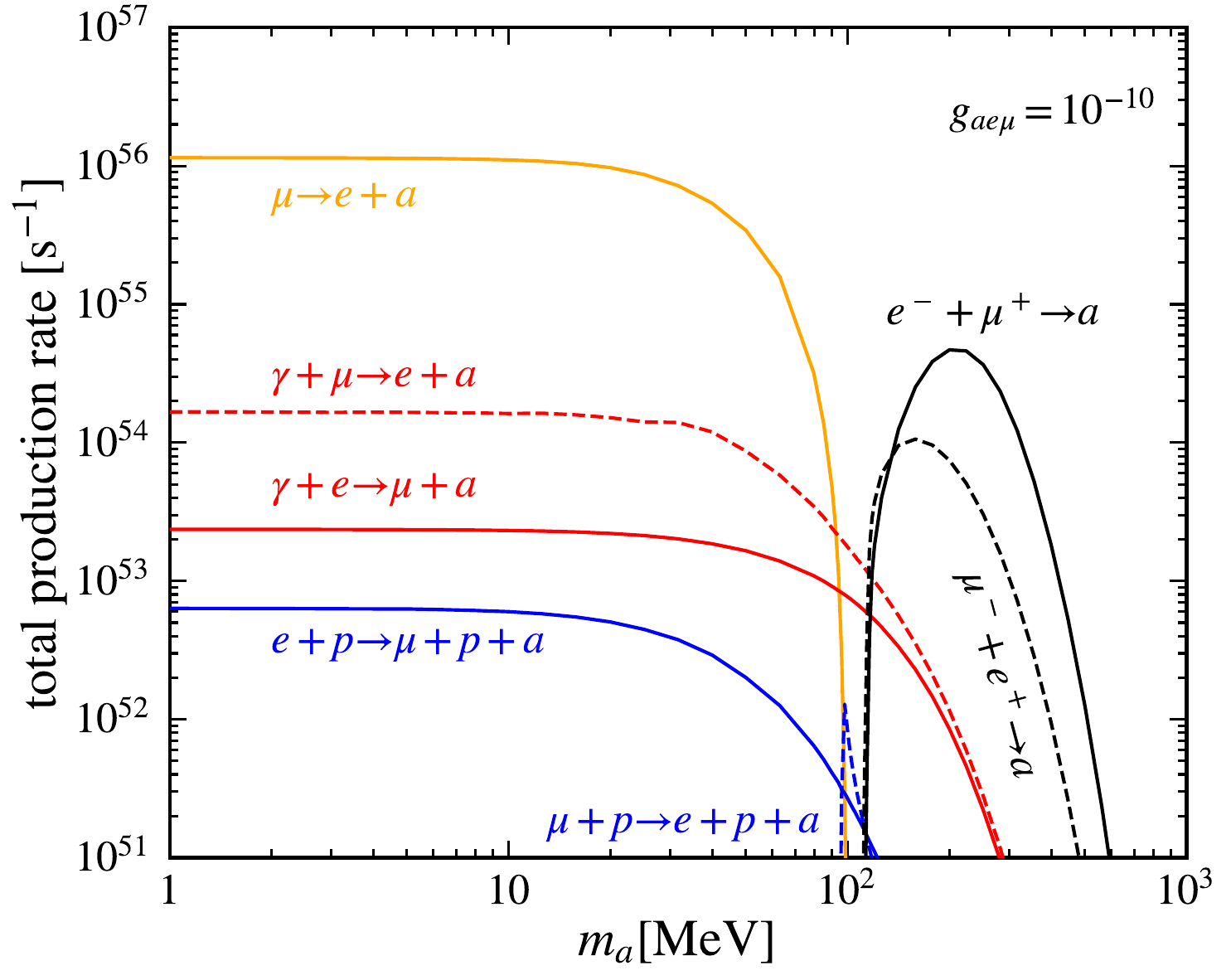}
    \caption{Total ALP production rate up to the gain radius 
    as a function of the ALP mass, for 
    different production channels: 
    (1) muon decay (orange), 
    (2) lepton bremsstrahlung: 
    $e+p\to \mu+p+a$ (blue solid) 
    and 
    $\mu+p\to e+p+a$ (blue dashed), 
    (3) $e$-$\mu$ coalescence: 
    $e^-+\mu^+\to a$ (black solid) 
    and $\mu^-+e^+ \to a$ (black dashed), 
    (4) semi-Compton: 
    $\gamma+e\to \mu+a$ (red solid) 
    and $\gamma+\mu\to e+a$ (red dashed). 
    Here we fix $g_{ae\mu} = 10^{-10}$. 
     }
    \label{fig:prod-compare}
\end{figure}

We compute the SN cooling constraints 
by computing the ALP luminosity 
via Eq.~\eqref{eq:lum-obs}, 
where we consider contributions from 
four ALP production channels: 
(i) muon decay, 
(ii) lepton bremsstrahlung, 
(iii) electron-muon coalescence, and
(iv) semi-Compton. 

{To compare different production channels, 
we compute the total production rate of ALPs 
within the gain radius via 
\begin{equation}
    \frac{dN_a}{dt} = 4\pi \int_0^{R_g} dr\ 
    r^2\int_{m_a'}^\infty dE_a\ \frac{d^2n_a}{dtdE_a}.  
\end{equation}
Fig.~(\ref{fig:prod-compare}) compares ${dN_a}/{dt}$ 
from the four different ALP production channels: 
For $m_a\lesssim 90$ MeV, 
the muon decay process is the dominant  
production channel. 
In the mass range of $m_a\gtrsim 115$ MeV, 
the electron-muon 
coalescence process is the dominant  
production channel. 
In the mass range 
of $m_a\sim(90,115)$ MeV, 
both the muon decay and electron-muon coalescence processes 
are kinematically forbidden, and 
the semi-Compton process provides the leading 
contributions to the ALP production. 
We find that the lepton bremsstrahlung processes 
are always subdominant.}

\begin{figure}[htbp]
\centering
\includegraphics[width=0.5\textwidth]{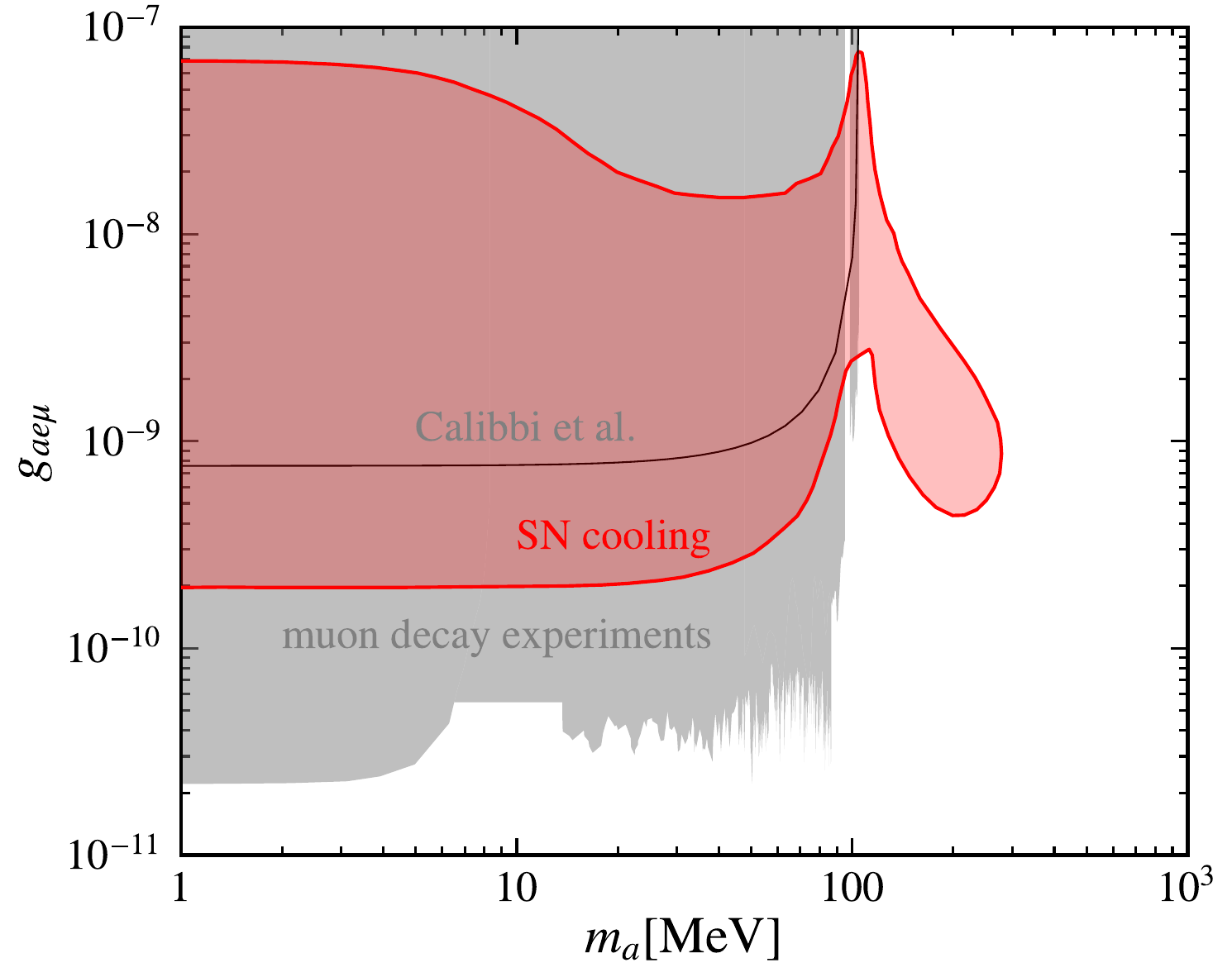}
\caption{The SN cooling constraints on LFV-ALPs (red), 
using the Garching muonic SN model SFHo-18.8 
\cite{garching-profile}.
We consider four LFV-ALP production processes: 
(i) muon decay, 
(ii) lepton bremsstrahlung, 
(iii) $e$-$\mu$ coalescence, and 
(iv) semi-Compton. 
Also shown is the SN cooling limit  
with the muon decay process only 
(black curve) \cite{Calibbi:2020jvd}. 
The gray shaded region shows the constraints from rare muon decay experiments: 
Derenzo \cite{Derenzo:1969za}, 
Bilger et al.\ \cite{Bilger:1998rp}, 
Jodidio et al.\ \cite{Jodidio:1986mz}, 
TWIST \cite{TWIST:2014ymv}, 
and PIENU \cite{PIENU:2020loi}.}
\label{fig:constraints}
\end{figure}

Fig.~(\ref{fig:constraints}) shows 
the SN cooling constraints on LFV-ALPs by 
demanding $L_a < L_\nu = 3\times 10^{52}$ erg/s, 
where the $L_a$ is the ALP luminosity. 
In the mass range of $m_a\sim (105,280)$ MeV,
the electron-muon coalescence process and 
semi-Compton process probe 
currently unexplored parameter regions.

We further compare our results with previous 
SN limits in Fig.~(\ref{fig:constraints}). 
The SN cooling limit from Ref.~\cite{Calibbi:2020jvd}, 
where only the muon decay process is considered, 
is approximately four times weaker than ours. 
This discrepancy is primarily due to 
the simple SN model used 
in Ref.~\cite{Calibbi:2020jvd}, 
where an SN core with constant density 
and temperature is assumed. 
In contrast, we adopt 
the Garching muonic SN model SFHo-18.8
\cite{Bollig:2020xdr,garching-profile}, 
which provides a more realistic treatment of the SN core and a more accurate determination of the muon abundance.

We also compare our results with constraints from rare muon decay 
experiments in Fig.~(\ref{fig:constraints}). 
We compute 
the branching ratio for the rare muon decay process: 
\begin{equation}
    {\rm Br}(\mu\to ea) = \frac{\Gamma(\mu\to ea)}{\Gamma(\mu\to e\nu\bar\nu)},
\end{equation}
where 
\cite{Calibbi:2020jvd}
\begin{equation}
    \Gamma(\mu\to ea) = \frac{g_{ae\mu}^2m_\mu}{16\pi}
    \left(1-\frac{m_a^2}{m_\mu^2}\right)^2, 
\end{equation} 
is the decay width of $\mu \to e + a$ 
with the electron mass neglected, 
and 
\begin{equation}
    \Gamma(\mu\to e\nu\bar\nu) = 
    \frac{G_F^2m_\mu^5}{192\pi^3}, 
\end{equation}
is the SM muon decay width 
\cite{ParticleDataGroup:2024cfk}, 
where $G_F=1.166\times 10^{-5}$ GeV$^{-2}$ 
is the Fermi constant. 
We then use the upper limit on 
${\rm Br}(\mu \to e + a)$ 
in Fig.~(1) of Ref.~\cite{PIENU:2020loi} to 
obtain the constraints, 
including  
the TWIST constraints for $m_a>45$ MeV, 
the constraints from   
Derenzo \cite{Derenzo:1969za}, 
from Bilger et al.\ \cite{Bilger:1998rp}, 
and from PIENU \cite{PIENU:2020loi}. 
For the constraints from rare muon experiments 
by Jodidio et al.\ \cite{Jodidio:1986mz} 
and TWIST \cite{TWIST:2014ymv} in the mass range of $m_a<45$ MeV, 
we adopt the limits 
from Fig.~(4) of Ref.~\cite{Zhang:2023vva}.

As shown in Fig.~(\ref{fig:constraints}), 
the constraints from rare muon decay experiments 
are about one order of magnitude stronger than  
the SN cooling constraints, 
in the mass range of $m_a\lesssim 105$ MeV. 
However, for $m_a\gtrsim 105$ MeV, 
the muon decay process is kinematically forbidden, 
and the only constraints shown in the parameter space 
of interest are the SN cooling constraints, 
which probe LFV-ALPs up to $m_a\sim 280$ MeV. 
We note that for $m_a > m_e + m_\mu$, other possible 
constraints could apply: the decay final states of 
long-lived ALPs produced in SNe 
(or binary neutron star mergers) 
could generate 
observable $X$-ray or $\gamma$-ray signals 
\cite{DelaTorreLuque:2023nhh,DelaTorreLuque:2023huu,
Balaji:2025alr}, or form fireballs, 
leading to sub-MeV photons \cite{Diamond:2023scc,Diamond:2023cto}.

Recently, Ref.~\cite{Fiorillo:2025yzf} has 
emphasized the importance of taking into account  
ALP production in the mantle for the net energy deposition in the 
mantle, in the strong coupling regime. 
Ref.~\cite{Fiorillo:2025yzf} also suggested that one should 
incorporate such effects when performing 
the SN cooling analysis. Indeed, 
ALPs produced in the mantle can be reabsorbed in the SN core, 
thereby 
reducing the net ALP luminosity that escapes the gain radius. 
To investigate such effects, we compute ALP 
luminosity for the following benchmark points: 
(a) $m_a = 105$ MeV and $g_{ae\mu}=10^{-7}$, and 
(b) $m_a = 200$ MeV and $g_{ae\mu} = 4\times10^{-9}$, 
which are slightly above the upper boundary of the 
exclusion region in Fig.~(\ref{fig:constraints}). 
For benchmark point (a), we find that 
$L_{\rm core} \simeq 4.7\times 10^{51}$ erg/s and 
$L_{\rm mantle} \simeq 8.5\times 10^{40}$ erg/s, 
where 
$L_{\rm core}$ denotes the luminosity from ALPs 
produced in the SN core and escaping the gain radius, and 
$L_{\rm mantle}$ denotes the luminosity from ALPs 
produced in the mantle and subsequently 
reabsorbed in the SN core. 
For benchmark point (b), 
$L_{\rm core} \simeq 8.8\times 10^{51}$ erg/s and 
$L_{\rm mantle} \simeq 4.6\times 10^{32}$ erg/s. 
Therefore, the effects due to ALP production in the mantle 
are insignificant for these two benchmark model points. 
We thus conclude that the upper boundary of the exclusion 
region in Fig.~(\ref{fig:constraints}) is 
essentially not affected 
by including ALP production in the mantle.

We note that, 
in the parameter space of interest, 
the ALP decay length is about $100-1000$ km, 
which is much smaller than the radius of the progenitor star. 
Consequently, ALPs are likely to decay in the SN mantle, 
leading to energy deposition that can 
contribute the SN explosion energy \cite{Caputo:2022mah}. 
Such effects allow one to constrain this parameter space 
using observations of low-energy SNe (LESNe) \cite{Huang:2025rmy, Huang:2025xvo}.

\section{Summary} 
\label{sec:sum}

In this paper, 
we compute the SN cooling constraints on 
LFV-ALPs 
that mediate $e$-$\mu$ lepton flavor violations.  
To properly compute the ALP luminosity, 
we consider the effects of ALP absorption  
in the SN. Moreover, 
we compute both ALP production and 
absorption terms up to the gain radius $R_g$. 
We note that our treatment consistently 
takes into account the ALP production between 
$R_\nu$ and $R_g$, which has been neglected 
in many previous studies.

We consider four LFV-ALP production channels in the SN core: 
(i) muon decay, 
(ii) lepton bremsstrahlung,  
(iii) $e$-$\mu$ coalescence, and
(iv) semi-Compton. 
The muon decay process is the dominant production channel 
in the low-mass regime, while the $e$-$\mu$ coalescence process 
is the dominant production channel 
at high masses. 
In the intermediate mass range, 
the semi-Compton process 
provides the leading contribution, 
dominating over the lepton bremsstrahlung process. 
We note that the $e$-$\mu$ coalescence  
and semi-Compton processes have not been 
considered previously in the literature for  
the SN cooling constraints on LFV-ALPs. 
We find that the two new channels
lead to the most stringent constraints 
on LFV-ALPs in the mass range of $\sim(105,280)$ MeV.

\acknowledgments

We thank 
Zi-Miao Huang, 
Changqian Li, Wenxi Lu and 
Zicheng Ye for discussions. 
We especially thank Hans-Thomas Janka 
for providing the SN profiles 
used for numerical calculations. 
The work is supported in part by the 
National Natural Science Foundation of China 
under Grant No.\ 12275128.  
The Feynman diagrams are created using 
the tikz-feynman package 
in \LaTeX~\cite{Ellis:2016jkw}.

\appendix

\begin{widetext}

\section{ALP production rate}
\label{sec:ALP-prod}

In this section we 
compute the ALP production rate. For this purpose, 
we consider a generic LFV-ALP production process $i\to j$, where 
$i$ and $j$ denote collectively the initial and final state particles, respectively, 
and one of the final state particles is an ALP. 
The other initial and final state particles are SM particles 
that are in the equilibrium state. In this case, 
the ALP production rate is given by 
(see e.g., 
Refs.~\cite{Kolb:1990vq,Raffelt:1990yz,Ferreira:2022xlw}) 
\begin{equation}
    \frac{d^2n_a}{dt dE_a} = \frac{|{\bf p}_a|}{4\pi^2}\int \prod_i d\Phi_i f_i\ \prod_{j\neq a} d\Phi_{j} (1\pm f_{j})(2\pi)^4\delta^{(4)}(P_i-P_j) \left|\mathcal{M}\right|^2, 
    \label{eq:prod-rate}
\end{equation} 
where $d\Phi_i = d^3p_i(2\pi)^{-3}(2E_i)^{-1}$ is the Lorentz 
invariant phase space 
for particle $i$ with momentum $P_i^\mu = (E_i, {\bf p}_i)$, 
$f_i$ is the equilibrium distribution function of particle $i$, 
$P_i$ ($P_j$)
is the total 
four-momentum of the initial (final) particles, and 
$\left|\mathcal{M}\right|^2$ is  
the squared matrix element with spin-sum for both initial and final states. 
We next compute the ALP 
production rates in the SN core for 
the following four processes: 
(i) muon decay,
(ii) lepton bremsstrahlung, 
(iii) $e$-$\mu$ coalescence, and
(iv) semi-Compton,
as shown in Fig.~(\ref{fig:feynman}).

\subsection{Muon decay}\label{sec:m-decay}

We first consider 
the muon decay process, $\mu^- \to e^- +a$, 
as shown in 
the upper left diagram of 
Fig.~(\ref{fig:feynman}). 
This process 
occurs when $m_\mu > m_a + m_e$. 
Because the effective electron mass is 
$\sim 10$ MeV in the SN core,
the muon decay process occurs for $m_a\lesssim 100$ MeV. 
The matrix element for the muon decay process is given by 
\begin{equation}
{\cal M}_d = -ig_{ae\mu} \bar{u}_e(P_2)\gamma_5 u_\mu(P_1), 
\label{eq:matrix-decay}
\end{equation} 
where $P_1$ and $P_2$ are the four-momenta of 
the $\mu^{-}$ and $e^-$ particles, respectively.
The ALP production rate
for the muon decay process is 
\begin{align}
    \frac{d^2n_d}{dtdE_a} =& \frac{|{\bf p}_a|}{4\pi^2}\int 
    d\Phi_1 d\Phi_2  f_\mu (1 - f_e)
    (2\pi)^4\delta^{(4)}(P_1-P_2-P_a)
    \left|{\cal M}_d\right|^2, 
\end{align}
where 
$f_e$ ($f_\mu$) is the Fermi-Dirac distribution for $e^-$ ($\mu^-$), and 
$\left|{\cal M}_{d}\right|^2= 
2g_{ae\mu}^2[(m_\mu-m_e)^2 - m_a^2]$ 
is the spin-summed matrix element. 
We first integrate out the phase space of the muon 
by using three of the four delta functions. 
Because $\left|{\cal M}_{d} \right|^2$ is constant, 
we can write 
$d^3p_2 = 2\pi {|{\bf p}_2|} E_2dE_2 d\cos\theta_{2a}$ 
so that 
\begin{align}
\frac{d^2n_d}{dtdE_a} =  \frac{\left|{\cal M}_{d} \right|^2}{32\pi^3} 
& \int \frac{|{\bf p}_2|}{E_1}dE_2 d\cos\theta_{2a}
f_\mu (1 - f_e) \delta(E_1-E_2-E_a).
\label{eq:nd-calc}
\end{align}
Noting that 
$E_1 = \sqrt{{m_1^2}+{\bf p}_2^2+{\bf p}_a^2+2|{\bf p}_2||{\bf p}_a|\cos\theta_{2a}}$, 
we integrate out $\cos\theta_{2a}$ to obtain 
\begin{equation}
    \frac{d^2n_d}{dtdE_a} = \frac{\left|{\cal M}_{d}\right|^2}{32\pi^3}\int_{E^-_2}^{E^+_2} dE_2\ f_\mu(1-f_e), 
    \label{eq:prod-rate-mdecay}
\end{equation}
where 
\begin{equation}
E^\pm_2 = \frac{E_a(m_1^2-m_2^2-m_a^2)}{2m_a^2}\pm \frac{\sqrt{E_a^2-m_a^2}I}{2m_a^2},
\end{equation}
where $I=\sqrt{(m_1^2-m_2^2-m_a^2)^2-4m_2^2m_a^2}$.

\subsection{Lepton bremsstrahlung}\label{sec:bremss}

We next consider the
lepton bremsstrahlung process, as shown 
in the upper-middle and upper-right diagrams of 
Fig.~(\ref{fig:feynman}). 
There are two different lepton bremsstrahlung processes: 
$e^-p\to \mu^- pa$ and $\mu^- p\to e^- pa$. 
Note that for the case where $m_\mu > m_a + m_e$, we consider the 
muon decay process of $\mu^- \to e^- +a$, 
but not the $\mu^- p\to e^- pa$ process; 
we consider the latter process only in the case 
where $m_\mu < m_a + m_e$.

For both $e^-p\to \mu^- pa$ and $\mu^- p\to e^- pa$, 
there are two diagrams 
(ALPs emitted either from the initial or from the final states), 
and the matrix element for each process 
is given by
\begin{equation}
    \begin{aligned}
         i&{\cal M}_{b} =- 
         4\pi\alpha 
         g_{ae\mu}[\bar{u}_4i\gamma^\nu u_2]\frac{ig_{\mu\nu}}{{(P_2-P_4)}^2}  \left[\bar{u}_3i\gamma^\mu 
         \frac{\slashed{P}+m_3}{P^2-m_3^2} 
         i\gamma_5 u_1+\bar{u}_3i\gamma_5 
         \frac{\slashed{Q}+m_1}{Q^2-m_1^2} 
         i\gamma^\mu u_1\right], 
    \end{aligned}
    \label{eq:bremss-mtrele}
\end{equation}
where 
$u_j \equiv u(P_j)$ is the Dirac spinor with 
$j=(1,2,3,4)$, 
$P = P_1 - P_a$, and 
$Q = P_3 + P_a$.
In this case, 
the ALP production rate is given by 
\begin{equation}
        \frac{d^2n_b}{dtdE_a} =\frac{|{\bf p}_a|}{4\pi^2}\int \prod_{i=1}^4 d\Phi_i f_1f_2(1-f_3)(1-f_4) 
        (2\pi)^4\delta^{(4)}(P_i-P_j)|{\cal M}_{b}|^2, 
\end{equation}
where $P_i=P_1+P_2$ and $P_j=P_3+P_4+P_a$.

To simplify the computation, we note that the proton mass 
is significantly larger than 
the temperature, $T\sim30$ MeV. 
We thus follow Ref.~\cite{Carenza:2021osu} to use the approximation 
where the proton is static such that $E_2\simeq E_4\simeq m_p$, 
which leads to
$\bar{u}_4^r\gamma^\nu u_2^s \simeq 2m_p \delta^{\nu 0} \delta^{rs}$ 
and $(P_2-P_4)^2 \simeq -|{\bf q}|^2$, 
where ${\bf q} = {\bf p}_2-{\bf p}_4 = {\bf p}_3+{\bf p}_a -{\bf p}_1$. 
We further include plasma effects on the photon propagator, 
such that the matrix element is 
\begin{align}
    i{\cal M}_{b} =2e^2 gm_p\frac{1}{|{\bf q}|\sqrt{{\bf q}^2+k_s^2}} 
    \left[\bar{u}_3\gamma^0\frac{\slashed{P}+m_3}{P^2-m_3^2}\gamma_5 u_1+\bar{u}_3\gamma_5\frac{\slashed{Q}+m_1}{Q^2-m_1^2}\gamma^0 u_1\right],
    \label{eq:bremss-mtrele-1}
\end{align}
where 
$k_s^2 = (4\pi\alpha/T) \sum_j Z_j^2n_j$ 
is the Debye screening scale 
with $n_j$ being the number 
density of ion $j$ with charge $Z_j e$ \cite{Raffelt:1985nk}; 
in our analysis, 
we only consider the contribution of protons.

We next integrate out $d\Phi_4$ by using of 
three of the four delta functions to obtain
\cite{Lucente:2021hbp,Ferreira:2022xlw}
\begin{align}
    \frac{d^2n_b}{dtdE_a} =&\frac{n_{\rm eff}}{8m_p^2}\frac{|{\bf p}_a|}{4\pi^2}\int d\Phi_1 d\Phi_3  f_1(1-f_3) 
    (2\pi)\delta(E_1-E_a-E_3)|{\cal M}_{b}|^2, 
\end{align}
where we have used the fact that the matrix element 
given in Eq.~\eqref{eq:bremss-mtrele-1} is independent of $p_2$, and
\begin{equation}
    n_{\rm eff} = 2\int\frac{d^3p_2}{(2\pi)^3}f_p(E_2) (1-f_p(E_4)), 
    \label{eq:neff}
\end{equation}
is the effective proton number density, where  
$f_p$ is the proton distribution function. 
Note that 
the Pauli blocking effects can be significant: 
including $f_p(E_4)$ as in Eq.~\eqref{eq:neff} can 
decrease $n_{\rm eff}$ up to 60\% in the SN core
 \cite{Ferreira:2022xlw}.
For protons 
in the SN core, we use 
\begin{equation}
    f_p(E_p) = \frac{1}{e^{(E_p-\mu_p)/T}+1},
    \label{eq:proton:occupation}
\end{equation} 
where 
$\mu_p$ is the proton chemical potential, and 
\begin{equation}
E_p = \sqrt{m_p^{*2}+{\bf p}_2^2} + U_p, 
\end{equation}
where 
$m_p^{*}$ is the effective proton mass, and 
$U_p$ is the proton interaction potential. 
In our analysis, we use $p_2 \simeq p_4$ 
and adopt the profiles of
$m_p^{*}$, $\mu_p$, and $U_p$ 
from the Garching profiles \cite{garching-profile}, 
which are shown in Fig.~(\ref{fig:SN-profile}).

The ALP production rate can be further simplified 
as follows \cite{Lucente:2021hbp,Ferreira:2022xlw}
\begin{align}
    \frac{d^2n_b}{dtdE_a} = 
    \frac{n_{\rm eff}|{\bf p}_a|}{(2\pi)^6 32m_p^2}\int_{m_3}^\infty dE_3\int_{-1}^{1}dz_adz_3 
    \int_0^{2\pi}d\phi\ |{\bf p}_1| |{\bf p}_3| f_1(1-f_3) \left|{\cal M}_b\right|^2,
    \label{eq:prod-rate-bremss}
\end{align}
where $|{\bf p}_1| = \sqrt{(E_3+E_a)^2-m_1^2}$, 
$\phi$ is the angle between 
the ${\bf p}_1$-${\bf p}_a$ plane and 
the ${\bf p}_1$-${\bf p}_3$ plane, 
$z_a=\cos\theta_{1a}$ with $\theta_{1a}$ being   
the angle between ${\bf p}_1$ 
and ${\bf p}_a$, and 
$z_3=\cos\theta_{13}$ with $\theta_{13}$ being 
the angle between ${\bf p}_1$ 
and ${\bf p}_3$.

\subsection{Electron-muon coalescence} 
\label{sec:emu-coa}

The electron-muon coalescence process of 
$e^\mp + \mu^\pm \to a$, as shown in 
the lower left diagram of Fig.~(\ref{fig:feynman}), 
occurs when $m_a > m_\mu + m_e$. 
We note that the $e^- + \mu^+ \to a$ process 
is more important than the $e^+ + \mu^- \to a$ process, 
primarily due to the higher distribution functions of 
the initial-state particles in the former case.
The matrix element for these two processes is 
${\cal M}_c = i g_{ae\mu}\bar{v}(P_2)\gamma_5u(P_1)$, 
where $P_1$ and $P_2$ are the four-momenta of the 
$e^-/\mu^-$ and $\mu^+/e^+$ particles, respectively.
The ALP production rate for this process is 
\begin{align}
    \frac{d^2n_{c}}{dtdE_a} = \frac{|{\bf p}_a|}{4\pi^2}\int d\Phi_1 d\Phi_2 (f_\mu f_e^+ + f_e f_\mu^+) 
    \times (2\pi)^4\delta^{(4)}(P_1+P_2-P_a)
    \left|{\cal M}_{c}\right|^2,
\end{align}
where $f_i^+ = \left[e^{(E_i+\mu_i)/T} + 1\right]^{-1}$ is the distribution of 
anti-fermions with $i=e,\mu$, and 
$\left|{\cal M}_{c}\right|^2 = 2g_{ae\mu}^2[m_a^2 - (m_\mu-m_e)^2]$ 
is the spin-summed matrix element.
Thus, we have
\begin{align}
    \frac{d^2n_{c}}{dtdE_a}  =& \frac{|{\bf p}_a|}{4\pi} \int \frac{d\Phi_2}{E_1}\ (f_\mu f_e^+ + f_e f_\mu^+) \delta(E_1+E_2-E_a)\left|{\cal M}_{c}\right|^2 \nonumber\\
    =& \frac{\left|{\cal M}_{c}\right|^2}{32\pi^3}\int_{E_2^-}^{E_2^+} dE_2\ (f_\mu f_e^+ + f_e f_\mu^+),
    \label{eq:prod-rate-coa}
\end{align}
where 
\begin{equation}
        E_2^\pm =\frac{E_a(m_a^2-m_1^2+m_2^2)}{2m_a^2} \pm \frac{\sqrt{E_a^2-m_a^2}}{2m_a^2}I.
\end{equation}

\subsection{Semi-Compton}
\label{sec:semi-compton}

Finally we consider the semi-Compton processes: 
$e+\gamma\to \mu+a$ and $\mu+\gamma\to e+a$. 
Each process proceeds via two channels, the $u$- and $s$-channel, 
shown in the lower-middle and lower-right diagrams of 
Fig.~(\ref{fig:feynman}), respectively. 
Recently, Ref.~\cite{Fiorillo:2025sln} found that 
the semi-Compton process dominates over the lepton bremsstrahlung process, 
for the $a$-$e$-$e$ coupling. 
While we confirm the qualitative conclusion 
of Ref.~\cite{Fiorillo:2025sln}, 
we also identify an error in their matrix element,  
which led to an overestimation of the axion production rate.

For both $\gamma (P_\gamma) + e (P_1) \to \mu (P_2) + a (P_a)$ 
and $\gamma  (P_\gamma) + \mu (P_1) \to e  (P_2) + a  (P_a)$, 
the matrix element is 
\begin{equation}
    i\mathcal{M}_{\rm Com} = e g_{ae\mu} \left[ \bar{u}_2(-i\gamma^\mu) 
    \epsilon_\mu\frac{i( 
    \slashed{P}_2
    -\slashed{P}_\gamma +m_2)}{u-m_2^2}\gamma_5u_1 + \bar{u}_2 \gamma_5 
    \frac{i(\slashed{P}_1 + \slashed{P}_\gamma +m_1)}{s-m_1^2}(-i\gamma^\mu)
    \epsilon_\mu u_1\right],
\end{equation}
where 
$P_\gamma$ is the four-momentum of the initial photon, 
$P_1$ ($P_2$) is the four-momentum of the initial (final) lepton, 
$m_1^2 = P_1^2$, 
$m_2^2 = P_2^2$, 
$s=(P_1+P_\gamma)^2$, and 
$u=(P_2-P_\gamma)^2$. 
The squared matrix element is thus
\begin{equation}
    \left|{\cal M}_{\rm Com}\right|^2 = 4\pi\alpha g_{ae\mu}^2 (\mathcal{F}_u 
    + \mathcal{F}_u + \mathcal{F}_{su}),
\end{equation}
where 
\begin{align}
    \mathcal{F}_u &= 4\frac{-(s-m_1^2)(u-m_2^2)+2(m_1m_2-m_2^2)(u-m_1m_2)+2m_2^2m_a^2}{(u-m_2^2)^2}, 
    \label{eq:fu}
    \\
    \mathcal{F}_s &= 4\frac{-(s-m_1^2)(u-m_2^2)+2(m_1m_2-m_1^2)(s-m_1m_2)+2m_1^2m_a^2}{(s-m_1^2)^2}, 
     \label{eq:fs}
     \\
    \mathcal{F}_{su} &= 8\frac{-(s-m_1m_2-m_a^2+m_2^2)(u-m_1m_2-m_a^2+m_1^2) + m_2^2 s + m_1^2 u - m_1^2m_2^2}{(u-m_2^2)(s-m_1^2)}.
    \label{eq:fsu}
\end{align}
Note that here we have neglected 
the plasma corrections on photons.
Also note that in the limit of $m_1\to 0$ and $m_2\to 0$, 
Eq.~\eqref{eq:fu} and Eq.~\eqref{eq:fs} agree with 
$\mathcal{F}_u$ and $\mathcal{F}_s$ 
in Ref.~\cite{Fiorillo:2025sln}, respectively; 
however, Eq.~\eqref{eq:fsu} differs from 
$\mathcal{F}_{su}$ 
in Ref.~\cite{Fiorillo:2025sln} by an overall sign. 
We find that the error in $\mathcal{F}_{su}$  
in Ref.~\cite{Fiorillo:2025sln} led to an 
overestimate of the axion production rate.

The ALP production rate in the semi-Compton process is given by 
\cite{Fiorillo:2025sln} 
\begin{align}
    \frac{d^2n_{\rm Com}}{dtdE_a} &= \int d\Phi_1 d\Phi_\gamma d\Phi_2 \frac{|{\bf p}_a|d\Omega_a}{2(2\pi)^3} f_1f_\gamma(1-f_2)(2\pi)^4\delta^{(4)}(P_1+P_\gamma-P_2-P_a)\left|{\cal M}_{\rm Com}\right|^2 \nonumber \\
    &= \int d\Phi_1d\Phi_\gamma\frac{1}{2E_2} \frac{|{\bf p}_a|d\Omega_a}{2(2\pi)^3} f_1f_\gamma(1-f_2)(2\pi)\delta(E_1+E_\gamma-E_2-E_a)\left|{\cal M}_{\rm Com}\right|^2,
    \label{eq:semicom-prod-rate}
\end{align}
where $f_\gamma(E_\gamma) = (e^{E_\gamma/T}-1)^{-1}$,  
$|{\bf p}_a|=\sqrt{E_a^2-m_a^2}$, and 
we have used the three delta functions for momentum conservation 
so that
$E_2 = \sqrt{({\bf p}_1 + {\bf p}_\gamma -{\bf p}_a)^2 +m_2^2}$ 
in the second line of Eq.~\eqref{eq:semicom-prod-rate}. 
The $\gamma +\mu\to e+ a$ process is divergent 
when the energy of the initial photon goes to zero. 
To remove such an IR divergence, 
we use the plasma frequency, $\omega_p$, 
as the lower cutoff on the photon energy, 
such that the dispersion relation of photons is given by
$E_\gamma^2 = |{\bf p}_\gamma|^2+\omega_p^2$. 
In the relativistic limit, the plasma frequency is 
given by 
\cite{Braaten:1993jw}
\begin{equation}
    \omega_p^2 = \frac{4\alpha}{3\pi}
    \left(\mu_e^2+\frac{1}{3}\pi^2T^2\right).
\end{equation}

Although Eq.~\eqref{eq:semicom-prod-rate} requires integration over eight variables, several of these variables do not enter 
the integrand (the matrix element and 
the delta function that ensures energy conservation) 
and are thus associated with integration symmetries, 
allowing them to be integrated out trivially. 
In our case, the matrix element is a function of $s$ and $u$, 
which can be expressed as 
\begin{align}
    &s = 2E_\gamma E_1-2|{\bf p}_\gamma||{\bf p}_1|X + m_1^2+\omega_p^2, 
    \\
    &u = m_1^2+m_a^2-2E_1E_a + \frac{2|{\bf p}_1||{\bf p}_a|}{|{\bf p}_{\rm tot}|}\left((|{\bf p}_1|+|{\bf p}_\gamma|X)z_a+|{\bf p}_\gamma|\sqrt{1-X^2}\sqrt{1-z_a^2}\cos\phi_{1a} \right), 
\end{align}
where 
$X \equiv \cos\theta_{1\gamma}$ with $\theta_{1\gamma}$ 
denoting the angle between ${\bf p}_1$ and ${\bf p}_\gamma$,
${\bf p}_{\rm tot}={\bf p}_1 + {\bf p}_\gamma$ such that 
$|{\bf p}_{\rm tot}|=\sqrt{{\bf p}_1^2+{\bf p}_\gamma^2+2|{\bf p}_1||{\bf p}_\gamma|X}$,
$z_a \equiv \cos\theta_a$ 
with $\theta_a$ denoting 
the angle between ${\bf p}_a$ and ${\bf p}_{\rm tot}$, 
and $\phi_{1a}$ is the angle between 
the ${\bf p}_{\rm tot}$-${\bf p}_1$ plane 
and 
the ${\bf p}_{\rm tot}$-${\bf p}_a$ plane. 
The energy-conservation delta function can be expressed as
\begin{equation}
\delta(E_1+E_\gamma-E_2-E_a) = \delta\left(E_1+E_\gamma-E_a-\sqrt{m_2^2 + {\bf p}_a^2+{\bf p}_{\rm tot}^2+2|{\bf p}_a||{\bf p}_{\rm tot}|z_a}\right).
\end{equation} 
Thus, we find that the integrand only depends on the 
following five variables: 
$|{\bf p_1}|$, $|{\bf p_\gamma}|$, $X$, $z_a$, and $\phi_{1a}$.

We next integrate out the integration variables 
that correspond to integration symmetries. 
Note that neither the orientation of the photon momentum 
(${\bf p}_\gamma$)
nor that of the initial lepton (${\bf p}_1$) 
appears in the integrand; 
only their relative angle ($X$) does. 
Thus, we can choose the $z$-axis along ${\bf p}_\gamma$ 
so that
\begin{equation}
d^3p_1 d^3p_\gamma = 8\pi^2 
{\bf p}_1^2 d|{\bf p}_1| 
dX {\bf p}_\gamma^2 d|{\bf p}_\gamma|, 
\end{equation}
where we have integrated out the solid angle of the photon 
and the azimuthal angle of ${\bf p}_1$, 
which are integration symmetries. 
Because the azimuthal angle of ${\bf p}_1$ is an 
integration symmetry, we can set it to be zero so that 
$\phi_{1a} = \phi_a$ and $d\Omega_a = dz_ad\phi_{1a}$. 
We then use the last delta function to remove the integration on 
$z_a$ in Eq.~\eqref{eq:semicom-prod-rate}: 
\begin{equation}
    \frac{d^2n_{\rm Com}}{dtdE_a} = \int_{m_1}^\infty |{\bf p}_1|dE_1\int_{\omega_p}^\infty |{\bf p}_\gamma| dE_\gamma \int_{-1}^1dX\int_0^{2\pi}d\phi_{1a} \frac{f_1f_\gamma(1-f_2)}{512\pi^6|{\bf p}_1+{\bf p}_\gamma|}\left|{\cal M}_{\rm Com}\right|^2\Theta\left(1-|z_a^0|\right),
    \label{eq:semicom-res}
\end{equation}
where  
\begin{equation}
    z_a^0 = \frac{2(E_1+E_\gamma)E_a-2E_1E_\gamma+2|{\bf p}_1||{\bf p}_\gamma| X-m_a^2-\omega_p^2-m_1^2+m_2^2}{2|{\bf p}_a||{\bf p}_{\rm tot}|}, 
\end{equation} 
and we have changed the integration variables from 
$d|{\bf p}_1|d|{\bf p}_\gamma|$ to $dE_1 dE_\gamma$. 
We note that in the limit of $m_1=m_2=0$ and $\omega_p=0$, 
Eq.~\eqref{eq:semicom-res} agrees 
Ref.~\cite{Fiorillo:2025sln}, 
which neglects both the electron mass and the plasma corrections to photons. The Heaviside function $\Theta$ in 
Eq.~\eqref{eq:semicom-res} 
selects the physical region 
of the parameter space: $|z_a^0|\leq 1$, 
which is equivalent to 
\begin{equation}
    X_{\rm min}\leq X\leq X_{\rm max},
\end{equation}
where $X_{\rm max/min} = \frac{1}{2|{\bf p}_a||{\bf p}_{\rm tot}|} 
\left[(|{\bf p}_a|\pm\sqrt{A})^2-|{\bf p}_1|^2-|{\bf p}_2|^2\right]$ 
with $A = (E_1+E_\gamma-E_a)^2-m_2^2$. Note that $A>0$ is always 
valid since $E_1+E_\gamma-E_a\geq m_2$ is guaranteed by the energy 
conservation. Thus, we have 
\begin{equation}
    \frac{d^2n_{\rm Com}}{dtdE_a} = \int_{m_1}^\infty |{\bf p}_1|dE_1\int_{E_{\rm\gamma min}}^\infty |{\bf p}_\gamma| dE_\gamma \int_{X_{\rm min}}^{X_{\rm max}}dX\int_0^{2\pi}d\phi_{1a} \frac{f_1f_\gamma(1-f_2)}{512\pi^6|{\bf p}_1+{\bf p}_\gamma|}\left|{\cal M}_{\rm Com}\right|^2\Theta\left(1-|X|\right),
    \label{eq:semicom-final}
\end{equation}
where $E_{\rm\gamma min} = \max\{\omega_p,\ E_a+m_2-E_1\}$. 
We use 
Eq.~\eqref{eq:semicom-final} 
in the numerical analysis for the semi-Compton process.

\section{The absorption of ALPs}\label{sec:abs}

In this section 
we calculate the absorption rate of LFV-ALPs 
in the SN. 
Consider a general ALP absorption process of $j\to i$, 
where $j$ and $i$ denote the initial and final 
state particles, respectively, 
and 
one of the initial state particles is an ALP $a$; 
all other 
particles in the $j\to i$ process 
are SM particles that are in equilibrium. 
The ALP absorption rate is given by 
\cite{Weldon:1983jn,Chang:2016ntp}
\begin{equation}
    \Gamma_{\rm abs} (E_a) = \frac{1}{2E_a}\int
    \prod_{j\neq a} d\Phi_j f_j \prod_i d\Phi_i (1\pm f_i)\ (2\pi)^4\delta^{(4)}(P_j-P_i)
    \left|{\cal M}\right|^2,
    \label{eq:abs-rate}
\end{equation}
where the quantities have the same definitions 
as in 
Eq.~\eqref{eq:prod-rate}. 
Note that 
in the 2-to-N process of $a+1 \to i$,  
Eq.~\eqref{eq:abs-rate} can be understood via 
$\Gamma_{\rm abs}^a = n_1 v \sigma(a+1 \to i)$, 
where $\sigma$ is the cross section, 
$v$ is the relative velocity between $a$ and $1$, and 
$n_1$ is the number density of particle $1$.

\begin{figure*}[htbp]
\centering
\includegraphics[width=0.28 \textwidth]{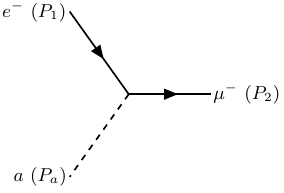}
\hspace{0.1cm}
\includegraphics[width=0.6 \textwidth]{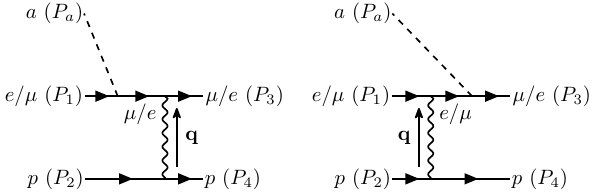}
\\[0.5 cm]
\includegraphics[width=0.28 \textwidth]{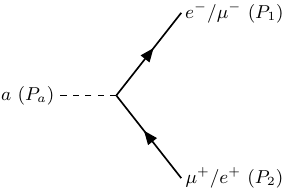}
\hspace{0.1cm}
\includegraphics[width=0.6\textwidth]{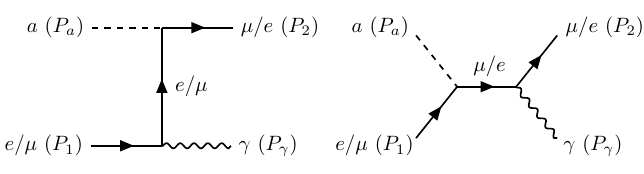}
\caption{
Absorption processes of LFV-ALP in the SN. Upper: $e$-$a$ coalescence (left) 
and inverse bremsstrahlung (middle and right). Lower: 
ALP decay (left) and inverse Compton (middle and right).
}
\label{fig:feynman-abs}
\end{figure*}

In our analysis, we consider 
ALP absorption through 
the following four processes: 
(1) $e$-$a$ coalescence, 
(2) inverse bremsstrahlung,  
(3) ALP decay, and 
(4) inverse-Compton, 
which are shown in Fig.~(\ref{fig:feynman-abs}); 
these correspond to  
the inverse processes of those shown in 
Fig.~(\ref{fig:feynman}).

The absorption rate given in Eq.~\eqref{eq:abs-rate} 
is related to the production rate given in  
Eq.~\eqref{eq:prod-rate}. 
First note that for both Fermi-Dirac and Bose-Einstein 
distributions, 
the distribution function $f$ 
obeys 
$1\pm f = e^{(E-\mu)/T} f$,
where the plus and minus signs 
on the left-hand side 
correspond to 
the Bose-Einstein and Fermi-Dirac distributions, respectively,  
and $E$ ($\mu$) is the energy (chemical potential) 
\cite{Weldon:1983jn}. 
This then leads to 
\begin{align}
    \Gamma_{\rm abs}(E_a) 
    &= \frac{1}{2E_a} \nonumber  \int \prod_{j\neq a} d\Phi_j e^{-(E_j-\mu_j)/T}(1\pm f_j) \prod_i d\Phi_i e^{(E_i-\mu_i)/T} f_i\ (2\pi)^4\delta^{(4)}(P_j-P_i) \left|{\cal M}\right|^2, \nonumber \\
    &= e^{(E_a-\mu_a^0)/T} \frac{1}{2E_a} \int \prod_i d\Phi_i f_i\ \prod_{j\neq a} d\Phi_j (1\pm f_j)(2\pi)^4\delta^{(4)}(P_j-P_i) \left|{\cal M}\right|^2, 
    \label{eq:prod-abs-A}
\end{align}
where 
we have used the energy conservation 
$E_a = \sum_i E_i - \sum_{j\neq a} E_j$ 
and defined 
$\mu_a^0 \equiv \sum_i\mu_i-\sum_{j\neq a} \mu_j$. 
By comparing Eq.~\eqref{eq:prod-abs-A} with 
Eq.~\eqref{eq:prod-rate}, one 
obtains the following relation between 
the absorption rate and the production rate: 
\begin{equation}
\Gamma_{\rm abs}(E_a) 
= e^{(E_a-\mu_a^0)/T} 
\frac{2\pi^2}{|{\bf p}_a|E_a}\frac{d^2n_a}{dt dE_a}.
\label{eq:prod-abs} 
\end{equation}
Thus the absorption rates of the four 
processes shown in 
Fig.~(\ref{fig:feynman-abs}) can be obtained 
through the production rates 
of their inverse processes, 
which are given in 
Eqs.~\eqref{eq:prod-rate-mdecay}, 
\eqref{eq:prod-rate-bremss}, 
\eqref{eq:prod-rate-coa}, and 
\eqref{eq:semicom-final}.

\end{widetext}

\normalem
\bibliography{ref.bib}{}
\bibliographystyle{utphys28mod}

\end{document}